\documentclass[article]{aa}  

\usepackage{graphicx}
\usepackage{txfonts}
\usepackage{hyperref}
\usepackage{natbib}
\bibpunct{(}{)}{;}{a}{}{,} 
\usepackage{color}
\usepackage{xspace}

\usepackage{listings}

\definecolor{codegreen}{rgb}{0,0.6,0}
\definecolor{codegray}{rgb}{0.5,0.5,0.5}
\definecolor{codepurple}{rgb}{0.58,0,0.82}
\definecolor{backcolour}{rgb}{0.95,0.95,0.92}

\lstdefinestyle{mystyle}{
    backgroundcolor=\color{backcolour},   
    commentstyle=\color{codegreen},
    keywordstyle=\color{magenta},
    numberstyle=\tiny\color{codegray},
    stringstyle=\color{codepurple},
    basicstyle=\ttfamily\footnotesize,
    breakatwhitespace=false,         
    breaklines=true,                 
    captionpos=b,                    
    keepspaces=true,                 
    numbers=left,                    
    numbersep=5pt,                  
    showspaces=false,                
    showstringspaces=false,
    showtabs=false,                  
    tabsize=2
}

\newcommand{\hipparcos}{{\it Hipparcos}\xspace}
\newcommand{\gaia}{{\it Gaia}\xspace}

\begin{document}

   \title{Characterizing and correcting the proper motion bias of the bright \gaia EDR3 sources}

    \titlerunning{The Bright Star \gaia EDR3 Reference Frame}
    \authorrunning{Cantat-Gaudin and Brandt}

   \author{Tristan Cantat-Gaudin
          \inst{1}
          \and
          Timothy D.~Brandt\inst{2}
          }

   \institute{
                Institut de Ci\`encies del Cosmos, Universitat de Barcelona (IEEC-UB), Mart\'i i Franqu\`es 1, E-08028 Barcelona, Spain
                \\
                 \email{tcantat@fqa.ub.edu}
         \and
         Department  of  Physics,  University  of  California,  Santa  Barbara, Santa Barbara, CA 93106, USA \\
             \email{tbrandt@ucsb.edu}
             }

   \date{}

 
  
  \abstract{In this paper we characterize magnitude-dependent systematics in the proper motions of the \gaia EDR3 catalog and provide a prescription for their removal.  The reference frame of bright stars ($G \lesssim 13$) in EDR3 is known to rotate with respect to extragalactic objects, but this rotation has proven difficult to characterize and correct.  
  We employ a sample of binary stars and a sample of open cluster members to characterize this proper motion bias as a magnitude-dependent spin of the reference frame. We show that the bias varies with $G$ magnitude, reaching up to 80\,$\mu$as\,yr$^{-1}$ for sources in the range $G$ = 11 -- 13, several times the formal EDR3 proper motion uncertainties. We also show evidence for an additional dependence on the color of the source, with a magnitude up to $\sim$10~$\mu$as\,yr$^{-1}$.  However, a color correction proportional to the effective wavenumber is unsatisfactory for very red or very blue stars and we do not recommend its use.  We provide a recipe for a magnitude-dependent correction to align the proper motion of the \textit{Gaia}~EDR3 sources brighter than $G=13$ with the International Celestial Reference Frame. }


   \maketitle
%

\section{Introduction}

The early third data release (EDR3) of the \gaia mission \citep{Gaia_EDR3} provides positions, proper motions, and parallaxes of more than 1 billion sources.  This astrometry is relative to the International Celestial Reference Frame \citep[ICRF,][]{Charlot20} defined by distant radio quasars.  
\gaia EDR3 spans more than 15 magnitudes in source brightness (more than a factor of $10^6$ in flux).  The \gaia satellite uses different readout modes (called window classes) and even different exposure times (implemented through gatings) to handle this huge dynamic range.  Different classes of observations must be calibrated to one another in order to align them with the ICRF.  This is particularly true for bright stars ($G \lesssim 13$), which have a different observation mode than the faint stars and for which there are no quasars of similar brightness.  

In \gaia DR2, the bright star reference frame rotated at $\sim$0.15\,mas\,yr$^{-1}$ relative to the faint stars and quasars \citep{Brandt18,Lindegren18}\footnote{We refer the reader to Appendix~B of \citet{Lindegren18} for further discussion of the origin of this calibration issue.}.
In EDR3, this rotation was tentatively removed by anchoring the window class 0 reference frame (WC0, corresponding to $G \lesssim 13$) to the positional difference of stars between the \hipparcos and \gaia missions \citep[see Sect.~4.5 of][]{LindegrenDR3astro}.
The \gaia bright star reference frame thus carries with it the 0.6\,mas uncertainty in the \hipparcos realization of the ICRF \citep{Kovalevsky+Lindegren+Perryman_etal_1997} divided by the $\approx$25-year interval between \hipparcos and \gaia.  
Throughout the magnitude range $G < 13$, \gaia uses a wide range of gatings; each transition carries with it the potential for systematic offsets in the reference frame.  \hipparcos provides almost no sources at magnitudes 11 -- 13 to serve as references.  

%

Here we use two independent sets of sources (resolved binaries and open clusters) to show that significant systematics are still present, and propose an additional correction to align the reference frame of the proper motions of the bright \textit{Gaia}~DR3 sources with that of the fainter sources and, by extension, the ICRF. Section~\ref{sec:data} presents the data used in this study. Section~\ref{sec:spin} investigates the dependence of the spin parameters on the magnitude and color of the sources. We conclude with Sect.~\ref{sec:conclusion}.

\section{Data used} \label{sec:data}

We calibrate the proper motions of the bright stars in \gaia EDR3 to those of the faint stars without referring to external data.  Our calibration sample consists of binaries with a bright ($G<14$) and faint ($G>14$) component, and of bound star clusters whose mean proper motions we estimate using their faint members.  
\gaia EDR3 calibrates the faint stars to quasars of similar magnitude.  
This section describes our samples of binaries and of open clusters.

\subsection{Binary stars} \label{sec:binaryselect}

The two components of a widely separated binary have similar motions through the Galaxy and nearly identical parallaxes. 
\cite{ElBadry+Rix+Heintz_2021} used these two facts to select pairs of stars in EDR3 whose parallaxes and proper motions are consistent with binarity.  Their catalog forms the basis of our binary star sample.
We seek to measure a frame rotation at the level of tens of $\mu$as\,yr$^{-1}$. 
For a pair of Solar mass stars at 500~pc, orbital motion is $\sim$100\,$\mu$as\,yr$^{-1}$ even at a separation of $10^4$~AU.  Our measurement of the frame rotation requires hundreds of widely separated binaries to average over this motion.

We select stars from the \cite{ElBadry+Rix+Heintz_2021} catalog using thresholds in angular separation and parallax, and requiring $1\,\mu{\rm m}^{-1} < \nu_{\rm eff} < 2\,\mu{\rm m}^{-1}$.  The latter criterion excludes six-parameter astrometric solutions \citep{Gaia_EDR3}.  We require stars to be separated by at least $5^{\prime\prime}$ to minimize the primary star's effect on the secondary's \gaia astrometry \citep{LindegrenDR3astro}. We then require that their rotational Keplerian velocity at their projected separation not exceed 0.2~mas\,yr$^{-1}$ assuming a system mass of 1~$M_\odot$.  The latter criterion is equivalent to
\begin{equation}
    2\pi \sqrt{\frac{\varpi^3}{\theta}} < 0.2~{\rm mas}
\end{equation}
with parallax $\varpi$ and angular separation $\theta$ both measured in mas.  Most binaries will have a non-zero eccentricity and will preferentially be observed at apastron.  Their physical separations will also be larger than their projected separations.  Both of these effects make the typical orbital motion a factor of a few smaller than our formal limit of 0.2~mas\,yr$^{-1}$.  Finally, we wish to select only true binaries, but spurious matches become increasingly common at wide angular separations.  We apply an additional cut that the chance alignment probability supplied by \cite{ElBadry+Rix+Heintz_2021} must be no higher than 5\%.  

The cuts above produce a sample of about 160,000 binaries.  We study the dependence of the reference frame on magnitude, anchoring the reference frame to that of the faint stars ($G \gtrsim 14$).  Of this sample of binaries, about 55,000 have a primary brighter than $G = 14$ and a secondary fainter than $G=14$.  These 55,000 stars, further divided by the magnitude of the primary, form the basis of the analysis in Sect.~\ref{sec:spin}.

\subsection{Open clusters}

Our open cluster sample uses the 1903 clusters listed in \citet{CantatGaudin20} that have members brighter than $G$ = 14. 
These clusters are distant and weakly bound, making them suitable calibrators for the EDR3 reference frame. 
We select stars from \gaia EDR3 within the central radius of each cluster, applying thresholds of 0.3\,mas in parallax and 0.5\,mas\,yr$^{-1}$ in proper motion around each cluster's listed centroids. 

Since \citet{CantatGaudin20} determined the clusters' mean astrometric parameters using \gaia~DR2, we recompute the median proper motion and parallax using our initial sample selected from EDR3.  We then repeat the query on EDR3 based on those updated values. After this second iteration we obtain about 358,000 stars, of which 37,000 are brighter than $G=14$.

\section{Aligning to the ICRF} \label{sec:spin}

\begin{figure*}
\begin{center}
\includegraphics[width=0.85\textwidth]{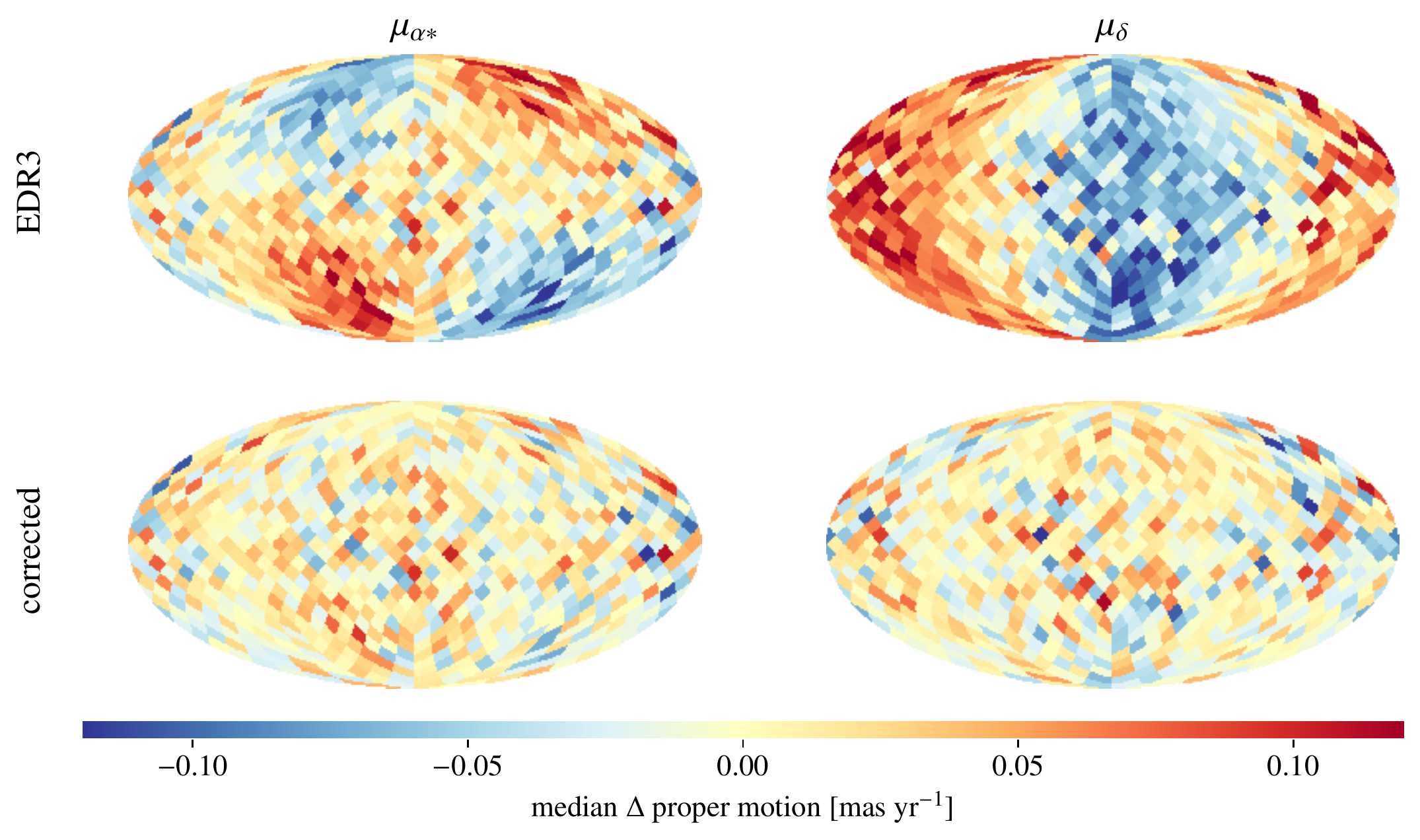}
\caption{ Median proper motion bias computed in healpix with Nside=8 for the sources in our sample in the magnitude range $11<G<13$, mapped in equatorial coordinates. The center is ($\alpha$,$\delta$)=(0,0), with $\alpha$ increasing toward the left.  
The lower panels show the results after the applying the correction derived in Sect.~\ref{sec:spin} using the prescription supplied in Appendix \ref{ap:python}.
\label{fig:mollweide}}
\end{center}
\end{figure*}

We begin by visualizing the proper motion bias across the sky using our combined sample of binaries and open cluster members.  We divide the sky into healpix tiles with ${\rm Nside} = 8$ and compute the median proper motion of stars in the magnitude range $11 < G < 13$ relative to the faint ($G>14$) stars.  The top panels of Fig.~\ref{fig:mollweide} show clear evidence for frame rotation, closely resembling that seen in DR2 \citep[c.f.~Fig~6 of][]{Brandt18}.  This likely reflects magnitude dependence of the anchoring of the window class 0 reference frame to the \hipparcos positions in 1991.25 \citep{ESA_1997,vanLeeuwen_2007}.  In this section, we derive the magnitude-dependence of this frame rotation and supply a prescription for its removal.  The lower panel of Fig.~\ref{fig:mollweide} shows the results of applying this correction to our full sample of $11 < G < 13$ stars.

\subsection{The Spin}
\label{subsec:spin}

\begin{figure}[!htp]
\begin{center}
\includegraphics[width=0.49\textwidth]{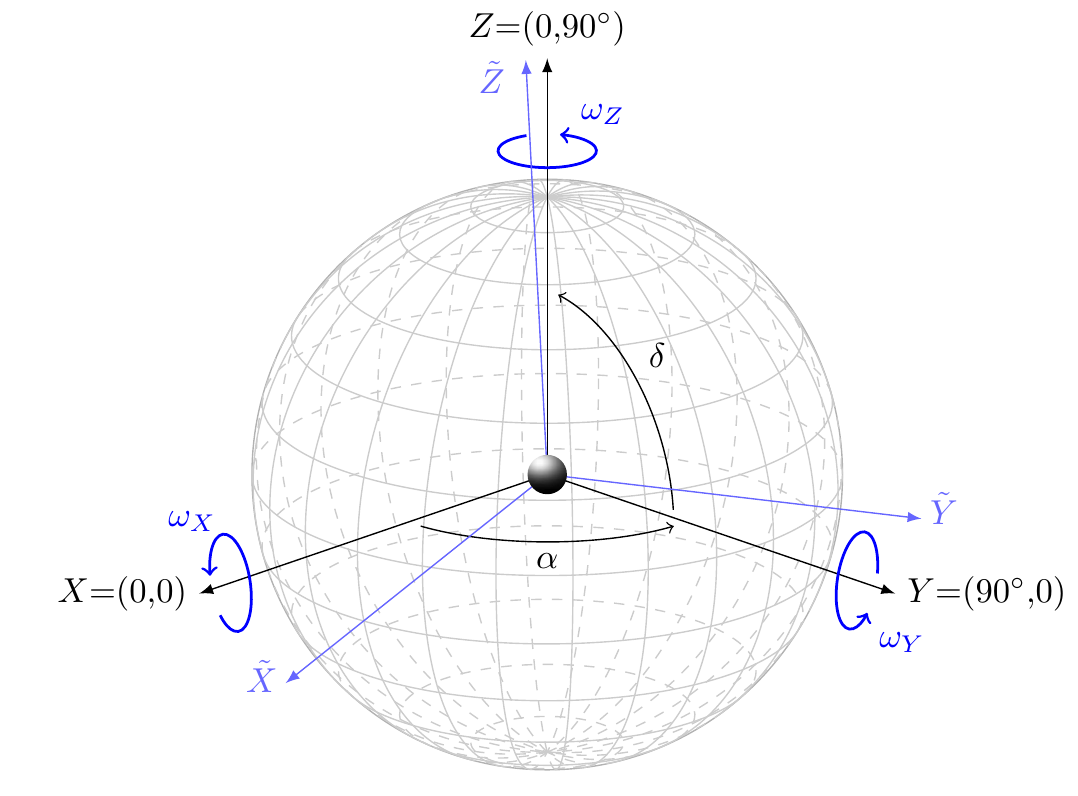}
\caption{ The ICRF is modeled as an orthogonal vector triad $\tens{C}=[\vec{X}~\vec{Y}~\vec{Z}]$. The transformation from to $\tens{C}$ to $\tens{\tilde{C}}=[\vec{\tilde{X}}~\vec{\tilde{Y}}~\vec{\tilde{Z}}]$ is obtained by applying the spin $\vec{\omega} = [\omega_X~\omega_Y~\omega_Z]$, which is a combination of rotations around the axes $\vec{X}$, $\vec{Y}$, and $\vec{Z}$.  
\label{fig:sphere}}
\end{center}
\end{figure}

We represent the ICRF as a vector triad $\tens{C}=[\vec{X}~\vec{Y}~\vec{Z}]$, where $\vec{X}$, $\vec{Y}$, and $\vec{Z}$ are orthogonal unit vectors pointing towards ($\alpha$,$\delta$)=(0,0), (90$^{\circ}$,0), and (0,90$^{\circ}$), respectively.
An arbitrary reference frame $\tens{\tilde{C}}=[\vec{\tilde{X}}~\vec{\tilde{Y}}~\vec{\tilde{Z}}]$ can be obtained by rotating $\tens{C}$ around the three axes $\vec{X}$, $\vec{Y}$, and $\vec{Z}$ as shown in Fig.~\ref{fig:sphere}.

Following the formalism presented in Sect.~2 of \citet{Lindegren20vlbi}, if the overall rotation $\vec{\omega} = [\omega_X~\omega_Y~\omega_Z]$ is small, the proper motion of a given source in the reference frame $\tens{\tilde{C}}$ is related to its proper motion in $\tens{C}$ through the approximation:
   \begin{equation}\label{eq:spindef}
     \begin{bmatrix}
       \tilde{\mu_{\alpha*}} - \mu_{\alpha*} \\
       \tilde{\mu_{\delta}}  - \mu_{\delta}  \\
     \end{bmatrix}
          =
     \begin{bmatrix}
       - \sin \delta \cos \alpha & - \sin \delta \sin \alpha & \cos \delta   \\
       \sin \alpha               & - \cos \alpha             & 0  \\
     \end{bmatrix}
     \begin{bmatrix}
       \omega_X \\
       \omega_Y \\
       \omega_Z
     \end{bmatrix}
   \end{equation}
The proper motion bias $(\Delta \mu_{\alpha*},\Delta\mu_\delta)$ introduced by the spin of the reference frame therefore depends on the $(\alpha,\delta)$ coordinates of the source. $\Delta\mu_\delta$ averages to zero over the whole sky, while $\Delta \mu_{\alpha*}$ is on average proportional to $\omega_Z$. 
This effect is illustrated in Fig.~\ref{fig:mollweide} for the stars with $11 < G < 13$ 
and is consistent with the findings of \citet{Fabricius20}, who point out a  low-significance offset of the $\mu_{\alpha*}$ of bright stars, and no detectable offset in $\mu_{\delta}$.

\subsection{Fitting the Spin Components} \label{sec:fit}

The proper motion difference between a bright source and its faint companion or cluster center is due to multiple factors. Part of the difference is due to measurement errors (assumed to be Gaussian) on the proper motions, which we denote as $(\sigma_{\alpha*,b},\sigma_{\delta,b}$) and $(\sigma_{\alpha*,f},\sigma_{\delta,f}$) for the bright and faint source, respectively.  For a cluster member, $\sigma_{\alpha*,f}$ and $\sigma_{\delta,f}$ are negligible because they represent medians over many stars.  An intrinsic velocity difference is also expected owing to the orbital motion of binaries, and to internal velocity dispersion in clusters.  We denote this intrinsic dispersion by $\sigma_1$. 

We also account for the fact that both our binary and cluster star samples contain a fraction $1-g$ of outliers that are background contaminants. Rather than performing e.g.~a sigma clipping, we model the proper motion distribution as a two-component Gaussian mixture by adding a broader component of dispersion $\sigma_2$.

The three spin components $\omega_X$, $\omega_Y$, and $\omega_Z$ give a proper motion shift $(\Delta \mu_{\alpha*}, \Delta \mu_\delta)$ at each star's position according to Eq.~\eqref{eq:spindef}.  We optimize these three components by maximizing the likelihood given by the Gaussian mixture model
\begin{align}
    {\cal L} = \prod_{{\rm stars}\,i} \Bigg( &\frac{g}{2\pi\sigma_{\alpha*,1}\sigma_{\delta,1}}  \exp\left[-\frac{(D\mu_{\alpha*})^2}{2\sigma_{\alpha*,1}^2} -\frac{(D\mu_{\delta})^2}{2\sigma_{\delta,1}^2} \right] \nonumber \\
    &+ \frac{1 - g}{2\pi\sigma_{\alpha*,2}\sigma_{\delta,2}}  \exp\left[-\frac{(D\mu_{\alpha*})^2}{2\sigma_{\alpha*,2}^2} -\frac{(D\mu_{\delta})^2}{2\sigma_{\delta,2}^2} \right] \Bigg) \label{eq:gaussmix}
\end{align}
with, e.g.,
\begin{equation}
\frac{(D\mu_{\alpha*})^2}{2\sigma_{\alpha*,1}^2} = \frac{(\mu_{\alpha*,b} - \mu_{\alpha*,f} - \Delta \mu_{\alpha*})^2}{2(\sigma_1^2 + \sigma_{\alpha*,b}^2 + \sigma_{\alpha*,f}^2)}.
\end{equation}

For the binary sample we take $\sigma_1$ = 100\,$\mu$as\,yr$^{-1}$, comparable to the typical orbital motion that we expect for these stars. For the clusters, $\sigma_1$ is the quadrature sum of 10\,$\mu$as\,yr$^{-1}$ (representing the effect of small-scale correlation on the proper motion precision) and an intrinsic proper motion dispersion corresponding to 0.5\,km\,s$^{-1}$ at the cluster distance. For both samples we use $g = \frac{1}{2}$ and $\sigma_2=0.3\,{\rm mas\,yr}^{-1}$. Varying $g$, $\sigma_1$ and $\sigma_2$ within reasonable values does not change our results, but affects the uncertainties by $\sim$10\%.


Maximizing Eq.~\eqref{eq:gaussmix} gives the best-fit frame rotation $\vec{\omega}$ via the proper motion shifts $(\Delta \mu_{\alpha*},\Delta\mu_\delta)$.  To assign uncertainties to these values, we use bootstrap resampling within each magnitude bin.  We report the mean and standard deviation of 400 bootstrap resamples for each magnitude bin.

\subsection{Spin parameters as a function of magnitude}

\begin{figure}
     \centering
     \includegraphics[width=\linewidth]{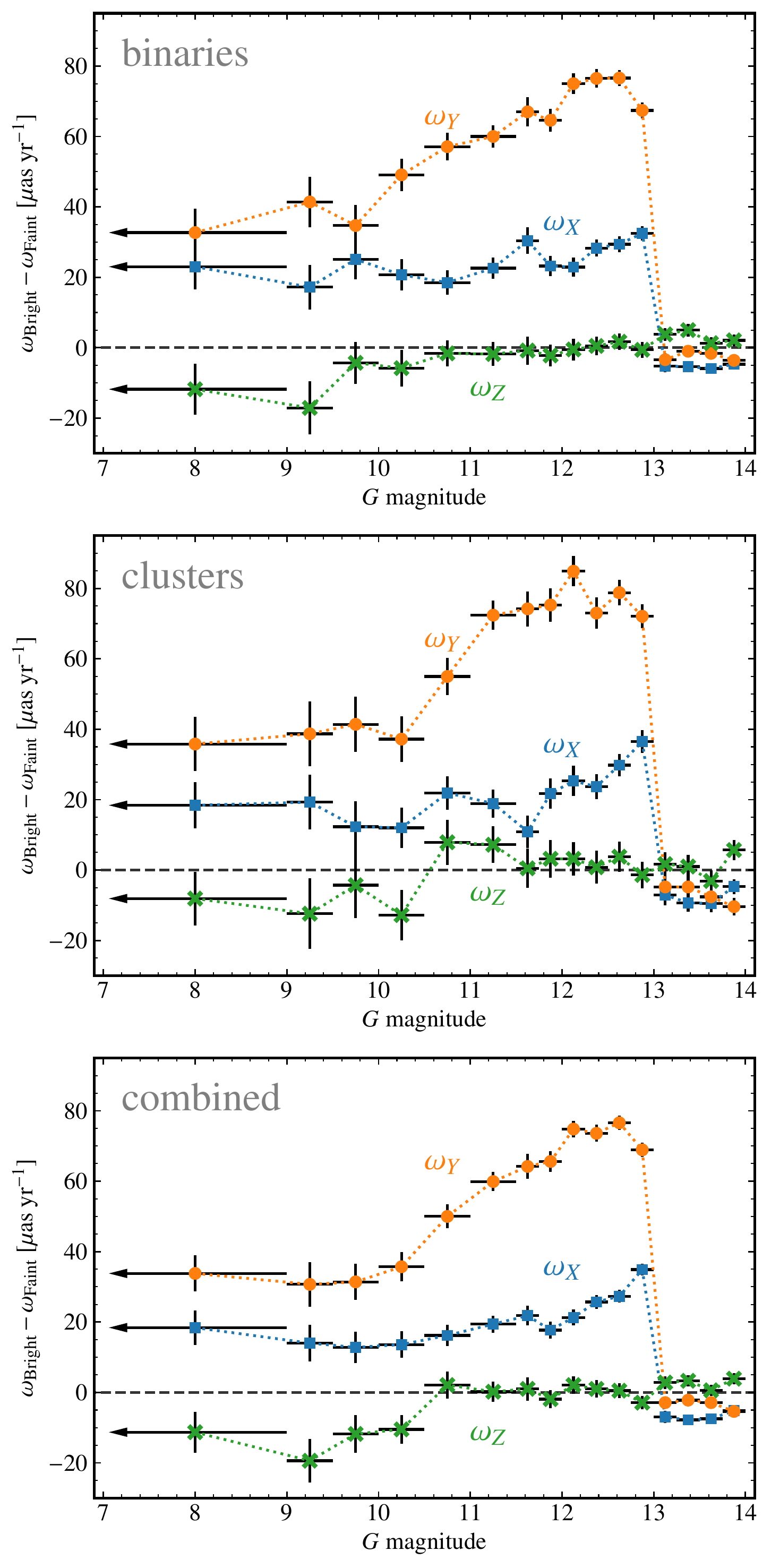}
     \caption{Components of the best-fit frame rotation derived from a sample of wide-separation binaries (top), open cluster members (middle), and both samples combined (bottom).  The uncertainties are derived from bootstrap resampling within each magnitude bin. \label{fig:spinParams} }
 \end{figure}

The best-fit spin components that we obtain in different magnitude bins are shown in Fig.~\ref{fig:spinParams} and listed in Table~\ref{table:omega}. 
The values obtained from binaries and from clusters are consistent, although the cluster sample provides slightly larger uncertainties due to a lower number of sources and to larger intrinsic velocity dispersions. The bottom panel of Fig.~\ref{fig:spinParams} shows the results obtained with the combined sample of binaries and cluster stars.

The $\omega_X$ and $\omega_Y$ components of the spin are very significant for the sources brighter than $G$ = 13, while $\omega_Z$ only appears significant at $G<10.5$. The $\omega_Y$ component varies strongly with magnitude, reaching up to 80\,$\mu$as\,yr$^{-1}$ for stars in the magnitude range $G = 12 - 13$. This bias is five times larger than the typical proper motion error quoted in the \textit{Gaia}~EDR3 catalog in this magnitude range.

\begin{figure*}
     \centering
     \includegraphics[width=\linewidth]{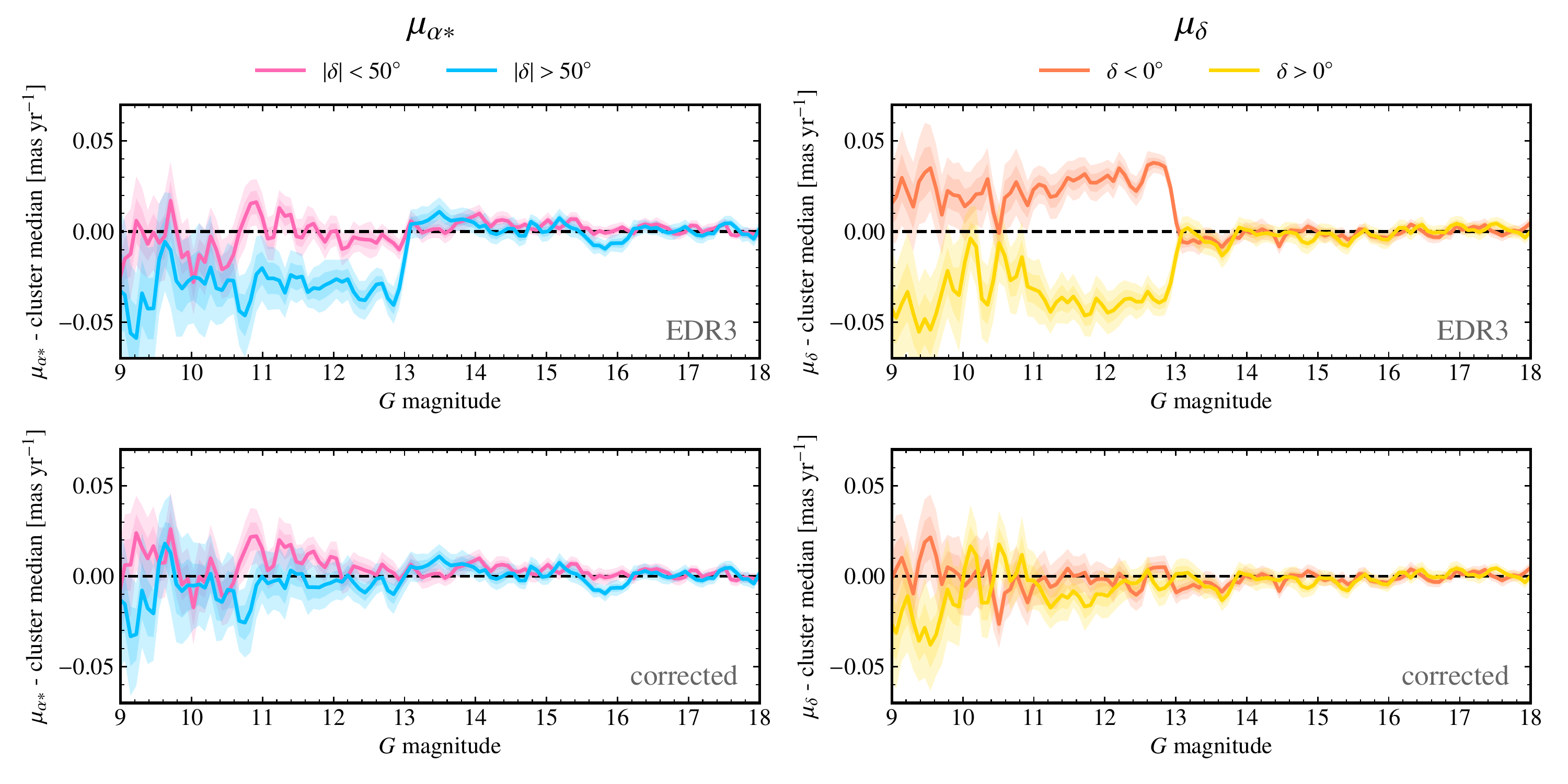}
     \caption{Top panels: running medians of the differences between proper motions of bright open cluster members and the proper motions of their host clusters as traced by the faint stars.  Bottom panels: the same running medians, after correcting the proper motions of the bright stars using the spin of the reference frame derived from binary stars.  
     The shaded areas correspond to 1-$\sigma$ and 2-$\sigma$ intervals computed using bootstrap resampling.    \label{fig:corrected} }
    \end{figure*}

Figure~\ref{fig:corrected} shows running medians of the difference between a star's proper motion and that of its cluster.  This bias depends on position; we divide the sky in half to avoid having it average to zero.  The median bias reaches $\sim$40~$\mu$as\,yr$^{-1}$ for $G \lesssim 13$, roughly double the formal EDR3 uncertainties, but would reach higher values if computed on smaller portions of the sky.
The bottom row of Fig.~\ref{fig:corrected} shows a correction of the proper motions of the cluster stars based on the spin computed from binary stars.
We align the proper motions of stars brighter than $G=13$ to the ICRF through the relation
   \begin{equation}\label{eq:correction}
       \begin{bmatrix}
        \mu_{\alpha*}  \\
        \mu_{\delta}   \\
    \end{bmatrix}_{ICRF}
    =
     \begin{bmatrix}
       \mu_{\alpha*} \\
       \mu_{\delta}  \\
     \end{bmatrix}_{EDR3}
          -
     \begin{bmatrix}
       - \sin \delta \cos \alpha & - \sin \delta \sin \alpha & \cos \delta   \\
       \sin \alpha               & - \cos \alpha             & 0  \\
     \end{bmatrix}
     \begin{bmatrix}
       \omega_X\\
       \omega_Y \\
       \omega_Z
     \end{bmatrix}
   \end{equation}
where the values of  $\omega_X$, $\omega_Y$, and $\omega_Z$ are the best-fit spin parameters obtained in the corresponding magnitude bin. We include in Appendix~\ref{ap:python} a Python function that performs the correction calibrated on the combined sample of cluster members and binaries.

For bright magnitudes ($G \lesssim 10$) the bias that we find likely reflects errors in the alignment of \hipparcos with the ICRF.  These were estimated to be 0.6~mas \citep{ESA_1997}, or about 25~$\mu$as\,yr$^{-1}$ if dividing by the difference between the \hipparcos and \gaia catalog epochs.  At $G=13$, the bias is similar to the rotation of $(\omega_X, \omega_Y, \omega_Z) = (-16.6, -95.0, +28.3)~\mu$as\,yr$^{-1}$ applied to bright sources \citep{LindegrenDR3astro}, with the signs reversed.  This suggests that the systematics in the bright source reference frame build gradually with magnitude away from the boundary between Window Classes 0 and 1.

    \begin{table}
\begin{center}
	\caption{ \label{table:omega} Spin best-fit parameters computed from the combined sample of binary stars and cluster members.}
	\begin{small}
\begin{tabular}{ll|ll|ll|ll}
\hline
\hline
  \multicolumn{1}{c}{$G_{min}$} &
  \multicolumn{1}{c}{$G_{max}$} &
  \multicolumn{2}{c}{$\omega_X$} &
  \multicolumn{2}{c}{$\omega_Y$} &
  \multicolumn{2}{c}{$\omega_Z$} \\
\hline
0.00 & 9.00 & 18.4 & $\pm$ 4.9 & 33.8 & $\pm$ 5.1 & -11.3 & $\pm$ 5.8  \\
9.00 & 9.50 & 14.0 & $\pm$ 5.2 & 30.7 & $\pm$ 6.3 & -19.4 & $\pm$ 6.2  \\
9.50 & 10.00 & 12.8 & $\pm$ 4.5 & 31.4 & $\pm$ 5.1 & -11.8 & $\pm$ 5.3  \\
10.00 & 10.50 & 13.6 & $\pm$ 3.8 & 35.7 & $\pm$ 4.2 & -10.5 & $\pm$ 4.1  \\
10.50 & 11.00 & 16.2 & $\pm$ 3.0 & 50.0 & $\pm$ 3.4 & 2.1 & $\pm$ 3.8  \\
11.00 & 11.50 & 19.4 & $\pm$ 2.4 & 59.9 & $\pm$ 2.7 & 0.2 & $\pm$ 2.8  \\
11.50 & 11.75 & 21.8 & $\pm$ 2.8 & 64.2 & $\pm$ 3.5 & 1.0 & $\pm$ 3.3  \\
11.75 & 12.00 & 17.7 & $\pm$ 2.4 & 65.6 & $\pm$ 3.0 & -1.9 & $\pm$ 2.6  \\
12.00 & 12.25 & 21.3 & $\pm$ 2.3 & 74.8 & $\pm$ 2.3 & 2.1 & $\pm$ 2.5  \\
12.25 & 12.50 & 25.7 & $\pm$ 2.0 & 73.6 & $\pm$ 2.4 & 1.0 & $\pm$ 2.5  \\
12.50 & 12.75 & 27.3 & $\pm$ 1.8 & 76.6 & $\pm$ 2.1 & 0.5 & $\pm$ 2.1  \\
12.75 & 13.00 & 34.9 & $\pm$ 1.7 & 68.9 & $\pm$ 2.1 & -2.9 & $\pm$ 1.9  \\
  \hline
\hline\end{tabular} 	\end{small}
\end{center}
\end{table}

\subsection{Evidence for a color dependence}

In order to calibrate color-dependent spread functions, the \gaia astrometric processing relies on prior color information given by the effective wavenumber $\nu_{\rm eff}$.  This is available in the EDR3 catalog for each source as \texttt{nu\_eff\_used\_in\_astrometry}. To investigate a color dependence of $\omega$ we allow the three spin parameters to vary linearly with $\nu_{\rm eff}$ by expressing them as
\begin{equation} \label{eq:cyyz}
    \begin{cases}
    \omega_X = \omega_{X,0} + c_X ( \nu_{\rm eff} - 1.5 ) \\
    \omega_Y = \omega_{Y,0} + c_Y ( \nu_{\rm eff} - 1.5 ) \\
    \omega_Z = \omega_{Z,0} + c_Z ( \nu_{\rm eff} - 1.5 ) \\
    \end{cases}
\end{equation}
This adds three free parameters to the fitting procedure described in Sect.~\ref{sec:fit}.

\begin{figure}
     \centering
     \includegraphics[width=\linewidth]{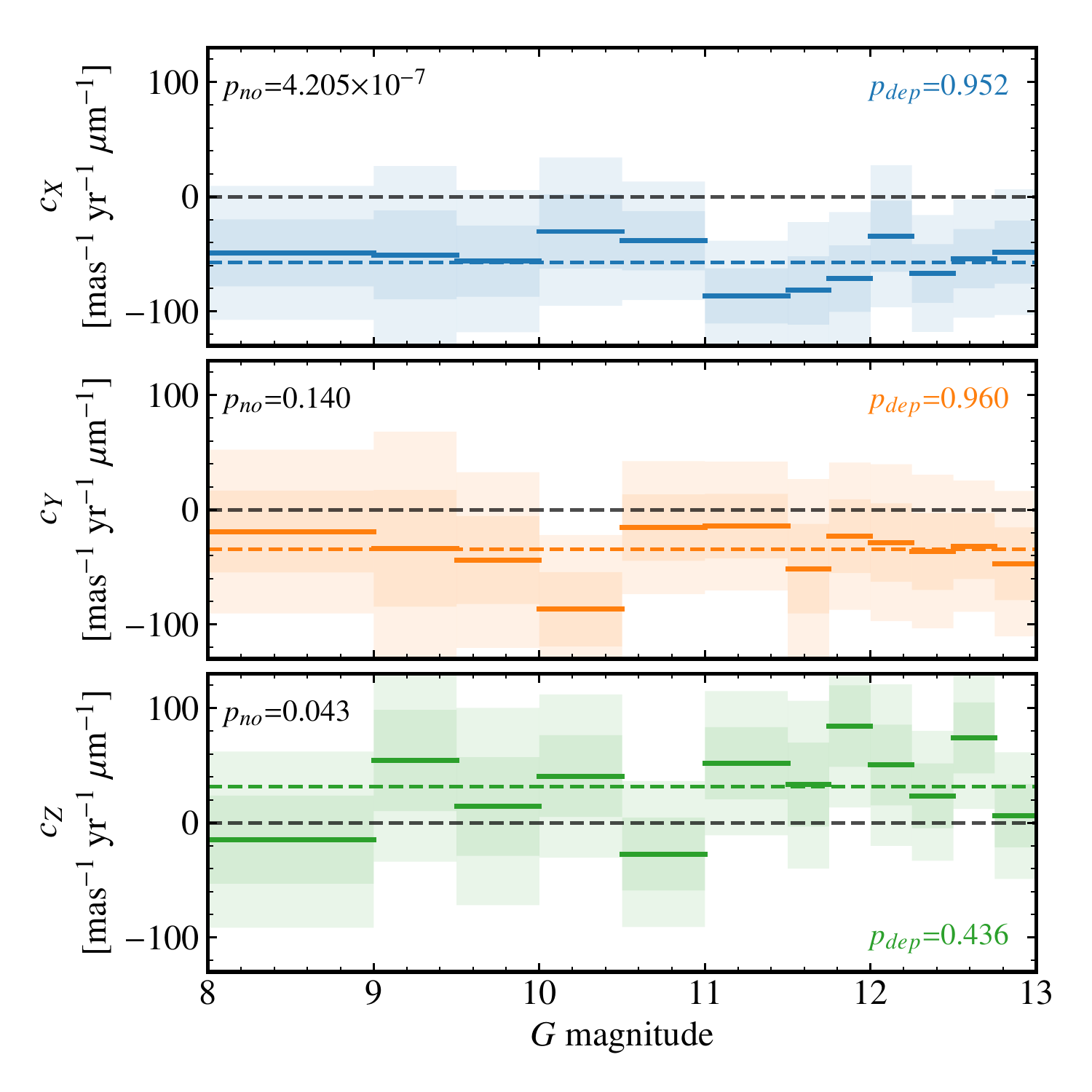}
     \caption{Coefficients obtained in different magnitude bins when fitting color-dependent spin parameters as defined in Eq.~\eqref{eq:cyyz}. The shaded areas correspond to 1-$\sigma$ and 2-$\sigma$ uncertainties obtained by bootstrapping. $p_{no}$ and $p_{dep}$ are defined in the text.  While $c_X$ in particular is significantly nonzero, we do not recommend the application of a color correction proportional to $\nu_{\rm eff}$.  Our sample lacks the statistical power to fully characterize the chromaticity of the spin.  \label{fig:cxyz} }
 \end{figure}

The best-fit values we obtain for these coefficients in each magnitude bin are shown in Fig.~\ref{fig:cxyz}, along with the uncertainty estimated by bootstrapping. We obtain negative values for $c_X$ and $c_Y$, indicating that the bias is stronger for smaller values of $\nu_{\rm eff}$ (corresponding to redder sources).  We find mostly positive values for $c_Z$.

To assess whether these values could be the result of statistical fluctuations, and whether $c_X$, $c_Y$, and $c_Z$ are constant with magnitude, we apply $\chi^2$ statistics.
In Fig.~\ref{fig:cxyz} we report the $p$-values corresponding to the hypothesis that the true value of $c_X$, $c_Y$, or $c_Z$ is zero ($p_{no}$), and to the hypothesis that they are non-zero and constant with $G$-magnitude ($p_{dep}$). We obtain small values for $p_{no}$, providing conclusive evidence that $\omega_X$ varies with $\nu_{\rm eff}$, and marginal evidence for $\omega_Y$ and $\omega_Z$. 
The $p_{dep}$ values indicate that, within the statistical precision allowed by our sample, $c_X$, $c_Y$, and $c_Z$ do not appear to vary with $G$ magnitude.
 
We next test whether this color correction, computed mainly from the binary stars, improves the proper motion residuals for the clusters members.  The binary stars and cluster members have systematically different $\nu_{\rm eff}$, with the bright cluster stars being more distant and bluer than the bright components of nearby, wide binaries.  The cluster stars also show a larger dispersion of $\nu_{\rm eff}$ values.  

We find that adding a color correction to the magnitude correction given by Eq.~\eqref{eq:correction} slightly increases the dispersion of the cluster stars' proper motion residuals, i.e., it degrades the quality of the correction.  We hypothesize that a linear dependence of the spin parameters on $\nu_{\rm eff}$ is a poor approximation for the wider range of colors represented by bright open cluster members: Eq.~\eqref{eq:correction} tends to overcorrect these stars.  Unfortunately, our sample is insufficient to fully characterize the dependence of the reference frame on color.  We therefore recommend only the magnitude correction given by Eq.~\eqref{eq:correction}, but advise that color-dependent frame rotations are present at $\sim$10~$\mu$as\,yr$^{-1}$ for a star somewhat bluer or redder than the \gaia's median of $\nu_{\rm eff} \approx 1.5$.

\section{Summary and conclusion} \label{sec:conclusion}
The large magnitude range covered by the \gaia observations requires the use of different gatings and readout modes. The reference frame of the faint sources can be calibrated directly to the International Celestial Reference Frame (ICRF) using distant radio quasars that have an optical counterpart. Due to the unprecedented astrometric capabilities of the \gaia spacecraft, no reference external sample of sufficient quality is available for brighter sources; the bright observations are calibrated to the fainter ones to align them with the ICRF.

In this paper we have shown that the reference frame of the proper motions of the bright EDR3 sources ($G<13$) rotates with respect to that of the faint EDR3 sources. The resulting proper motion bias reaches 80\,$\mu$as\,yr$^{-1}$ in the magnitude range $G=$ 11 -- 13, five times larger than the nominal uncertainty listed in the \gaia EDR3 catalog for these sources.
We have also shown evidence for a second-order dependence of the spin parameters on the color of the source when modeled as a linear function of the effective wavenumber $\nu_{\rm eff}$. 
A color correction proportional to $\nu_{\rm eff}$, however, proves unsatisfactory over the wide range of colors present in our open cluster sample.  We recommend the use of only a magnitude-dependent correction, and caution that color-dependent systematics of up to $\sim$10~$\mu$as\,yr$^{-1}$ will still be present.

Our main result is a magnitude-based correction to be applied to the \gaia~EDR3 proper motions of bright stars in order to align them with the ICRF.  The correction may be computed using Eq.~\eqref{eq:correction} with the spin parameters listed in Table~\ref{table:omega}.  Appendix~\ref{ap:python} provides sample Python code to realize this correction for a list of stars in \gaia EDR3.

\section*{Acknowledgments}
We thank L. Lindegren, C. Fabricius, and F. Arenou for their feedback and helpful comments.
TCG acknowledges support by the Spanish Ministry of Science, Innovation and University (MICIU/FEDER, UE) through grants RTI2018-095076-B-C21 and the Institute of Cosmos Sciences University of Barcelona (ICCUB, Unidad de Excelencia `Mar\'{\i}a de Maeztu’) through grant CEX2019-000918-M. 
This work has made use of data from the European Space Agency (ESA) mission Gaia (https://www.cosmos.esa.int/gaia), processed by the Gaia Data Processing and Analysis Consortium (DPAC; https://www.cosmos.esa.int/web/gaia/dpac/consortium). Funding for the DPAC has been provided by national institutions, in particular the institutions participating in the Gaia Multilateral Agreement.

\bibliographystyle{aa}
\bibliography{refs}

\appendix
\onecolumn
\section{Python implementation of the frame spin correction} \label{ap:python}

This function applies Eq.~\eqref{eq:correction} using the values listed in Table~\ref{table:omega}.

\begin{lstlisting}[language=Python,numbers=none]
def edr3ToICRF(pmra,pmdec,ra,dec,G):
    """
    Input: source position, coordinates,
           and G magnitude from Gaia EDR3.
    Output: corrected proper motion.
    """
    if G>=13:
        return pmra, pmdec

    import numpy as np

    def sind(x):
        return np.sin(np.radians(x))
        
    def cosd(x):
        return np.cos(np.radians(x))

    table1=""" 0.0   9.0     18.4    33.8   -11.3
               9.0   9.5     14.0    30.7   -19.4
               9.5   10.0    12.8    31.4   -11.8
               10.0  10.5    13.6    35.7   -10.5
               10.5  11.0    16.2    50.0    2.1  
               11.0  11.5    19.4    59.9    0.2  
               11.5  11.75   21.8    64.2    1.0  
               11.75 12.0    17.7    65.6    -1.9 
               12.0  12.25   21.3    74.8    2.1  
               12.25 12.5    25.7    73.6    1.0  
               12.5  12.75   27.3    76.6    0.5  
               12.75 13.0    34.9    68.9    -2.9 """
    table1 = np.fromstring(table1,sep=' ').reshape((12,5)).T

    Gmin = table1[0]
    Gmax = table1[1]

    #pick the appropriate omegaXYZ for the source's magnitude:
    omegaX = table1[2][(Gmin<=G)&(Gmax>G)][0]
    omegaY = table1[3][(Gmin<=G)&(Gmax>G)][0]
    omegaZ = table1[4][(Gmin<=G)&(Gmax>G)][0]

    pmraCorr = -1*sind(dec)*cosd(ra)*omegaX -sind(dec)*sind(ra)*omegaY + cosd(dec)*omegaZ
    pmdecCorr = sind(ra)*omegaX -cosd(ra)*omegaY

    return pmra-pmraCorr/1000., pmdec-pmdecCorr/1000.


\end{lstlisting}

\end{document}